\documentstyle[12pt,twoside,fleqn,espcrc1,epsf]{article}

\newcommand{\be}{\begin{equation}}
\newcommand{\ee}{\end{equation}}
\newcommand{\bea}{\begin{eqnarray}}
\newcommand{\eea}{\end{eqnarray}}
\newcommand{\pardis}{\langle \mu \rangle}

\newcommand{\AmS}{{\protect\the\textfont2
  A\kern-.1667em\lower.5ex\hbox{M}\kern-.125emS}}

\title{Color confinement and dual superconductivity in unquenched QCD
}

\author{
    J.M. Carmona\address[ZARA]{Departamento de F\'{\i}sica Te\'orica, 
    Universidad de Zaragoza, 50007 Zaragoza, Spain },
    M. D'Elia\address[GENO]{Dipartimento di Fisica dell'Universit{\`a} 
    and INFN, I-16146, Genova, Italy},
    L. Del Debbio\address[PISA]{Dipartimento di Fisica  
    and INFN, I-56127 Pisa, Italy},
    A. Di Giacomo\addressmark[PISA] \thanks{Speaker at the 
    Conference (Email: digiaco@df.unipi.it). Partially supported by MIUR and 
    by EC, FMRX-CT97-0122}, 
    B. Lucini\address[OXFO]{Theoretical Physics, University of Oxford,
    1 Keble Road, OX1 3NP Oxford, UK},
    G. Paffuti\addressmark[PISA], C. Pica\addressmark[PISA]}

\begin{document}

\maketitle

\begin{abstract}
We report on evidence from lattice simulations that confinement is produced 
by dual superconductivity of the vacuum in full QCD as in quenched QCD.
Preliminary information is obtained on the order of the deconfining phase 
transition.
\end{abstract}

\section{Introduction}
\label{sec:introduction}

Confinement of color and the deconfining phase transition have been observed
in numerical simulations on the lattice.
In pure gauge theory, (SU(2) and SU(3)), the transition is well understood.
$\langle L \rangle$, the expectation value of the Polyakov line, is 
a good order parameter and $\langle \tilde{L} \rangle$, the expectation
value of the 't Hooft line, is a good disorder parameter~\cite{vort}; 
the corresponding symmetries are $Z_N$, $\tilde{Z}_N$.

The situation is less clear in the presence of dynamical quarks (full QCD).
There the symmetry $Z_N$ ($\tilde{Z}_N$) is explicitely broken by the coupling
of the quarks. In the chiral limit $m_q = 0$ the chiral parameter 
$\langle \bar{\psi} \psi \rangle$ would be a good order parameter for
the chiral phase transition; however at $m_q \neq 0$ also chiral symmetry 
is explicitely broken. Moreover it is not clear a priori what relation exists 
between chiral symmetry and confinement.

For a model QCD with two quarks ($N_f = 2$) of equal mass $m_u = m_d = m_q$
the phase diagram is schematically represented in Fig.~1. The line of 
phase transition in the plane $m_q$, $T$ is defined by the maxima of 
susceptibilities~\cite{karsch,jlqcd}, among which 
\be
\chi_L = \int d^3 x \langle L(\vec{x},0) L^\dagger(\vec{0},0) \rangle
\ee
and the susceptibility $\chi_{\rm ch}$ of the chiral order parameter
\be
\chi_{\rm ch} = \int d^3 x \langle \bar{\psi}\psi(\vec{x},0) \bar{\psi}\psi (\vec{0},0) \rangle \; .
\ee
All of them have a maximum at the same value of $T$, for a given $m_q$,
which defines the line in Fig.~1.
For $m_q >$ 3 GeV, the maxima of $\chi_L$ diverge proportionally to the 
volume $V$,
indicating a first order transition. At $m_q = 0$ there are theoretical 
reasons and numerical indications that the transition is 
second order~\cite{piwi,karsch,jlqcd}.
At intermediate values of $m_q$
the susceptibilities studied in~\cite{karsch,jlqcd} stay finite as 
$V \to \infty$, and this is interpreted as a crossover.

In the usual approach to phase transitions the free energy (effective
lagrangean in the language of field theory) depends on the order 
parameters in a form which is dictated by symmetry and by scale analysis.
The susceptibilities which keep memory of the order of the transition
are those of the order parameters. At intermediate quark masses
neither $\langle L \rangle$ nor  $\langle \bar{\psi} \psi \rangle$
are good order parameters.

\begin{figure}[htp]
\label{diagram}
\vspace{-4pt}
\begin{center}
\leavevmode
\epsfxsize=76mm
\epsffile{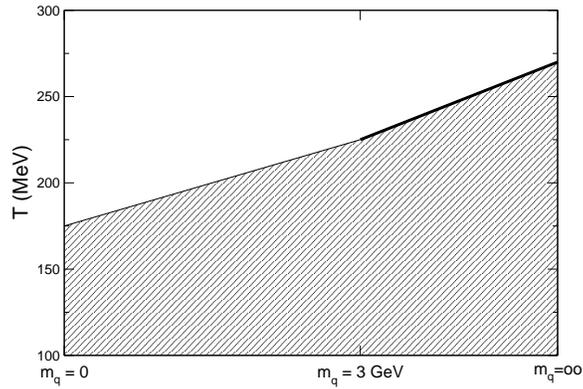}
\end{center}
\vspace{-35pt}
\caption{Phase diagram for two degenerate flavours, $m_u = m_d = m_q$.}
\vspace{-40pt}
\end{figure}
\begin{figure}[htp]
\label{highbeta}
\vspace{-4pt}
\begin{center}
\leavevmode
\epsfxsize=70mm
\epsffile{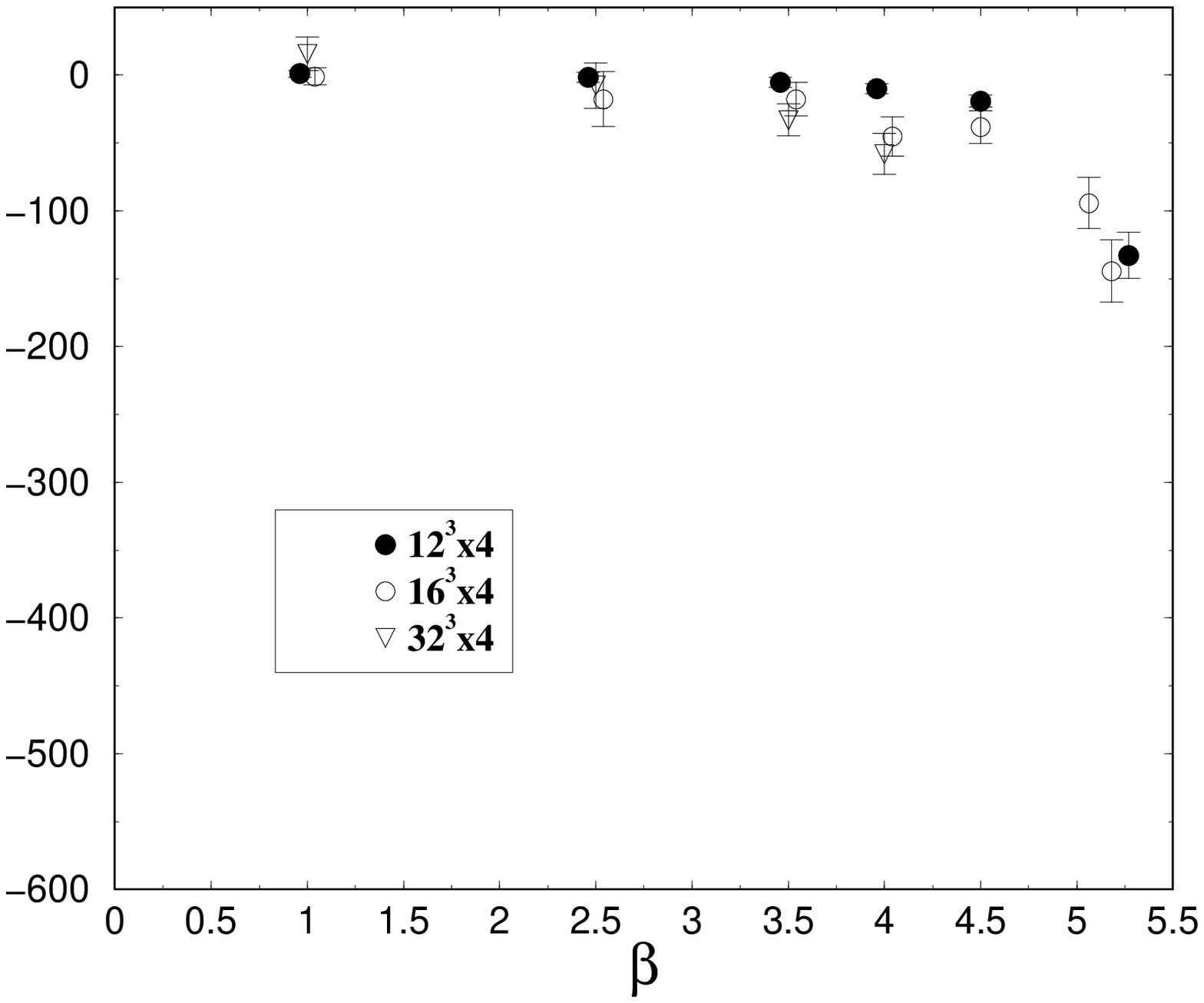}
\epsfxsize=70mm
\epsffile{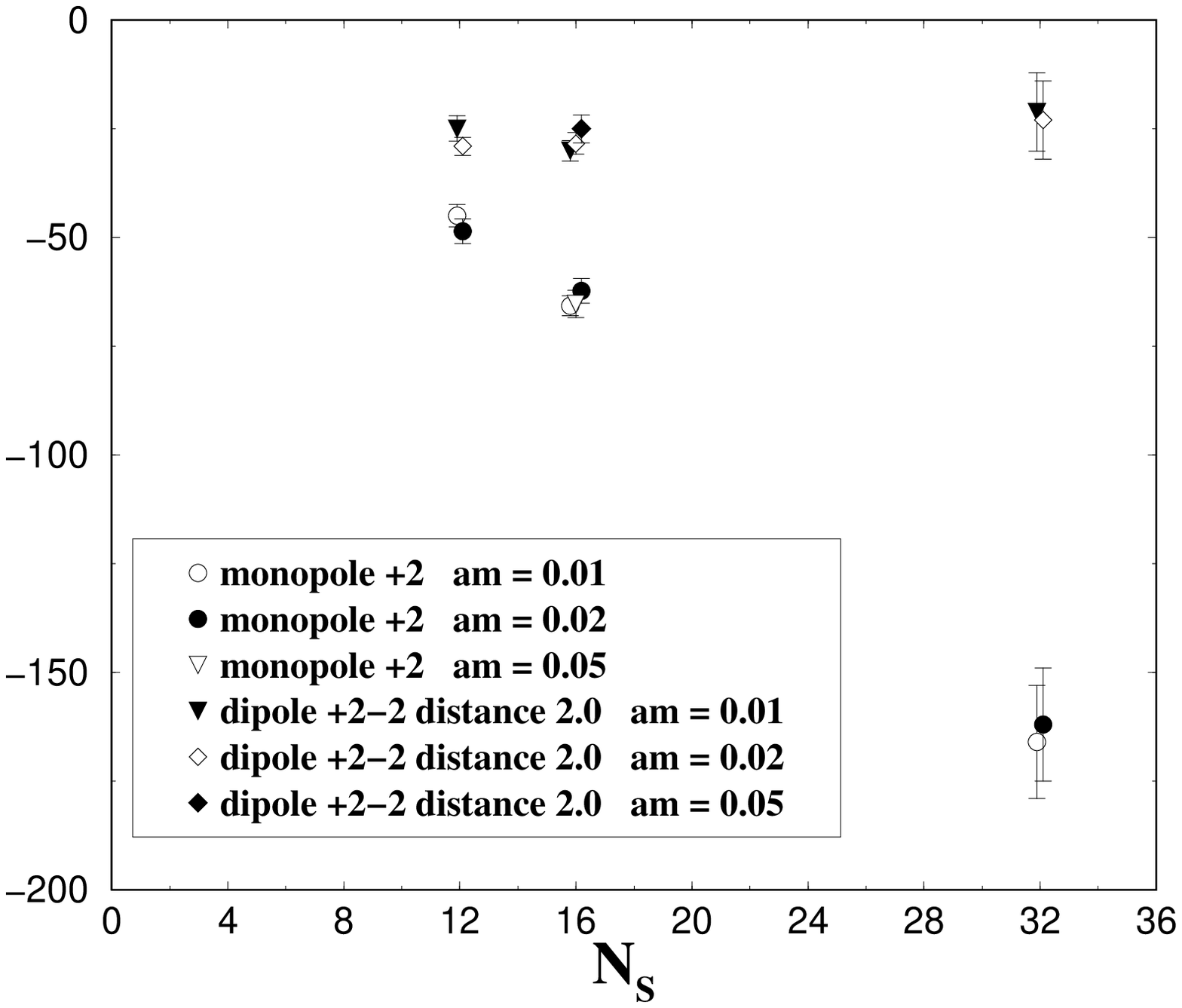}
\end{center}
\vspace{-35pt}
\caption{$\rho$ as a function of $N_s$ for $T < T_c$ (left-hand
side). $\rho$ as a function of $N_s$ for $T > T_c$: as $N_s \to
\infty$, $\rho \to - \infty$ if the magnetic charge is non-zero and
stays constant otherwise (right-hand side).}
\vspace{-18pt}
\end{figure}

A good order parameter exists, which is the vacuum expectation value
of a magnetically charged operator $\mu$. $\pardis \neq 0$ signals dual
superconductivity of the vacuum~\cite{ldd,I,II,III}. According 
to the ideas of ref.'s~\cite{'tHooft75,Mandelstam76}, dual superconductivity channels 
the chromoelectric field acting between a $q \bar{q}$ pair into an Abrikosov
flux tube, with energy proportional to the length, $V = \sigma r$.

In quenched QCD~\cite{I,II,III} a detailed analysis of $\pardis$ shows
that the vacuum is indeed a dual superconductor for $T < T_c$
($\pardis \neq 0$), and makes a transition to normal at $T_c$. Above
$T_c$, $\pardis = 0$. As $T \to T_{c-}$, 
$\pardis \simeq \tau^\delta \phi (a/\xi,\xi/N_s)$, 
with $\tau = (1 - T/T_c)$, $\xi \sim \tau^{-\nu}$ the physical 
correlation length and $N_s$ the linear spatial extension of the lattice.
At the critical point $a/\xi \ll 1$ and can be put to zero, while the variable
$\xi/N_s$ can be traded with $N_s^{1/\nu} \tau$.
As a consequence the scaling law 
$\pardis = \tau^\delta \tilde{\phi} (N_s^{1/\nu} \tau)$ holds.
A better variable than $\pardis$ is 
$\rho = \frac{\partial}{\partial \beta} \log \langle \mu \rangle$,
which is a susceptibility. In terms of $\rho$,
\be
\pardis = \exp \left( \int_0^\beta \rho(\beta') d \beta ' \right)
\label{mubyrho}
\ee The scaling law $\rho/N_s^{1/\nu} = f(\tau N_s^{1/\nu})$ allows
one to control the infinite volume limit $N_s \to \infty$, which is
necessary to have a phase transition, and to extract $\nu$, $T_c$,
$\delta$.  For quenched QCD, $\pardis$ is equally good as $\langle
\tilde{L} \rangle$ and is numerically coincident with it~\cite{vort}.

\begin{figure}[t]
\label{mpiconst}
\vspace{-4pt}
\begin{center}
\leavevmode
\epsfxsize=76mm
\epsffile{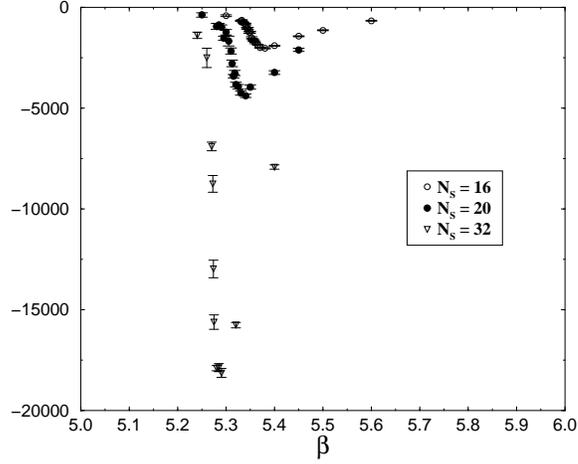}
\end{center}
\vspace{-35pt}
\caption{$\rho$ peak at different values of $N_s$. The values of $m_q$
are chosen such that $m_q N_s^\gamma$ is kept constant 
($m_q = 0.075$ at $N_s = 16$,
 $m_q = 0.043$ at $N_s = 20$ and $m_q = 0.01335$ at $N_s = 32$).}
\vspace{-18pt}
\end{figure}

A similar analysis can be repeated for full QCD~\cite{IV}.
$\pardis$ is defined as in quenched  and defines the same symmetry.
Simulations have been done on $N_t \times N_s^3$ lattices with $N_t = 4$
and $N_s = 12,16,20,32$. The time extension $N_t$ determines the transition
temperature, the space extension $N_s$ is used to perform the finite size 
scaling analysis of the infinite volume limit.
The first result is that for $T < T_c$, $\pardis \neq 0$ as $N_s \to \infty$.
Indeed in that range of temperature, $\rho$ is $N_s$ independent (see
Fig.~2).
For $T > T_c$ instead $\rho \simeq -k N_s + c$ ($k > 0$) as 
$N_s \to \infty$ (see Fig.~2), i.e., by Eq.~(\ref{mubyrho}),
$\pardis$ is strictly zero in the infinite volume limit.

\begin{figure}[ht]
\label{rhopeak}
\vspace{-4pt}
\begin{center}
\leavevmode
\epsfxsize=79mm
\epsffile{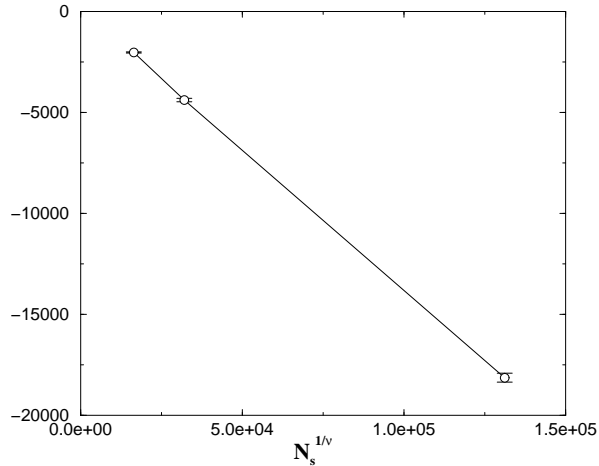}
\end{center}
\vspace{-35pt}
\caption{Values of $\rho$ at the peak as a function of $N_s^{1/\nu}$, with
$\nu = 1/3$. The height of the peak clearly scales as $N_s^{1/\nu}$.}
\vspace{-18pt}
\end{figure}

Around $T_c$, again,
\be
\pardis = 
\tau^\delta \phi\left( \frac{a}{\xi},\frac{N_s}{\xi},m_q N_s^\gamma \right) \; .
\label{scaling}
\ee
A new scale, $m_q$, appears which was absent in the quenched case. 
However the exponent $\gamma$ is known ($\gamma \simeq 2.49$~\cite{karsch,jlqcd}) and simulations can be made choosing $m_q$ and $N_s$ such that
$m_q N_s^\gamma$ in Eq.~(\ref{scaling}) is kept fixed. 
Then, neglecting $a/\xi$ again, the scaling law becomes
\be
\pardis = 
\tau^\delta \phi\left( 0,\frac{N_s}{\xi},{\rm const.} \right)  \; ,
\; \rho/N_s^{1/\nu} = f(\tau N_s^{1/\nu}) \; .
\label{scaling2}
\ee 
The behaviour of $\rho$ around $T_c$ is shown in Fig.~3 for different
values of $N_s$: it has a peak at $T_c$, Eq.~(\ref{scaling2}) implies
in particular that the values of $\rho$ at the peak ($\tau = 0$)
$\bar{\rho}$, scale as $\bar{\rho} \propto N_s^{1/\nu}$, whence $\nu$
can be determined. The result is shown in Fig.~4: contrary to the
common belief, the transition is first order. This result is still
preliminary and will be cross-checked by more simulations, and by the
use of improved actions. A scenario in which confinement is controled
by one single order parameter $\pardis$ and by one and the same
symmetry pattern (dual superconductivity) is appealing and meets the
requirements of $N_c \to \infty$ arguments.

\end{document}